\documentclass[cits]{PoS}

\usepackage{amsmath}
\usepackage{amssymb}
\usepackage{graphicx}
\usepackage{dsfont}
\usepackage{dcolumn}
\usepackage{units}
\usepackage{bm}
\usepackage{wasysym}
\usepackage{multirow}
\usepackage{slashed}
\usepackage{setspace}

      \newcommand{\be}{\begin{equation}}
      \newcommand{\ee}{\end{equation}}

      \newcommand{\Slash}[1]{\slashed{#1}}

\title{Nucleon Compton scattering from the Dyson-Schwinger perspective}

\ShortTitle{Nucleon Compton scattering}

\author{\speaker{Gernot Eichmann}  \\
        Institut f\"ur Physik, Karl-Franzens-Universit\"at Graz, 8010 Graz, Austria\\
        E-mail: \email{gernot.eichmann@uni-graz.at}}

\author{Christian S. Fischer\\
        Institut f\"{u}r Theoretische Physik, Justus-Liebig-Universit\"at Giessen, D-35392 Giessen, Germany\\
        E-mail: \email{christian.fischer@theo.physik.uni-giessen.de}}

\abstract{We summarize results from our recent paper\thanks{G. Eichmann and C. S. Fischer, arXiv:1212.1761 [hep-ph]. } ~on     
          nucleon Compton scattering in the Dyson-Schwinger approach.
          We study the Compton scattering phase space and its various kinematic limits, and we introduce a covariant, transverse tensor basis for the scattering amplitude that is
          free of kinematic singularities. We illustrate the decomposition of the Compton amplitude
          at the quark-gluon level and calculate its nonperturbative handbag part, which also includes all $t-$channel meson resonances.
          The calculation requires numerical solutions for the dressed quark propagator, the nucleon's Faddeev amplitude and the quark Compton vertex as its input;
          they are consistently obtained from Dyson-Schwinger, Bethe-Salpeter and Faddeev equations.
          As a first application, we verify that the result for the Compton scattering amplitude reproduces
          the $\pi\gamma\gamma$ transition form factor when the pion pole in the $t$ channel is approached.}

\FullConference{Xth Quark Confinement and the Hadron Spectrum,\\
		October 8-12, 2012\\
		TUM Campus Garching, Munich, Germany}

\begin{document}

  \section{Introduction}

     \begin{figure*}[t]
     \center{
     \includegraphics[scale=0.102]{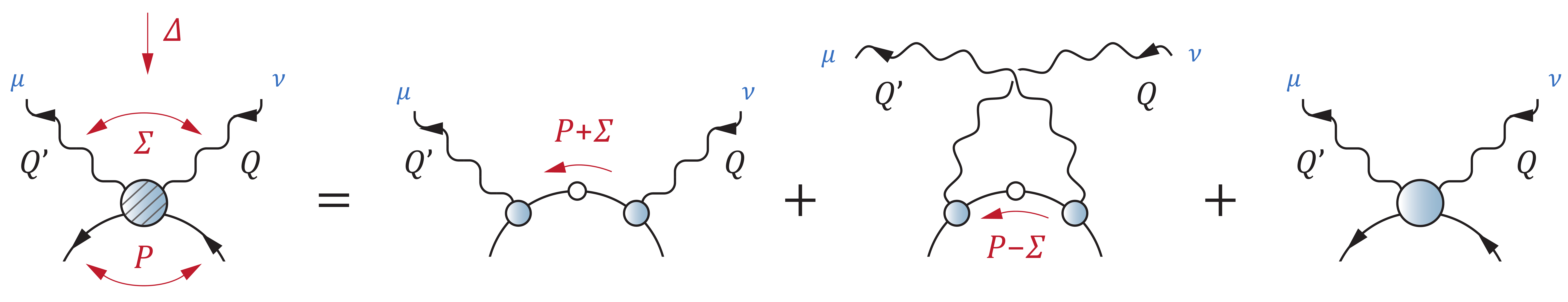}}
        \caption{Separation of the Compton amplitude into Born terms and a structure part. }
        \label{fig:qcv-born}
     \end{figure*}

   \renewcommand{\arraystretch}{1.0}

     Nucleon Compton scattering is a powerful tool for probing the structure of the
     nucleon and is pursued at various experimental facilities such as MAMI, JLab, MIT-Bates and HI$\gamma$S at Duke.
     In the low-energy region, the Compton amplitude encodes the nucleon's (generalized) polarizabilities that describe the nucleon's response to an applied electromagnetic field.
     At high energies, Compton scattering probes the fundamental quark and gluon degrees of freedom via the nucleon's generalized parton distributions.
     The integrated Compton amplitude contributes to two-photon exchange processes,
     and its imaginary part in the forward limit is linked to the 
     nucleon structure functions via the optical theorem.
     There are many excellent reviews on the topic from experimental and different theoretical
     perspectives~\cite{Belitsky:2001ns,Drechsel:2002ar,HydeWright:2004gh,Schumacher:2005an,Drechsel:2007sq,Downie:2011mm,Pasquini:2011zz,Griesshammer:2012we}.

             In Ref.~\cite{Eichmann:2012mp} we have started to investigate Compton scattering from
             the Dyson-Schwinger approach~\cite{Alkofer:2000wg,Fischer:2006ub,Roberts:2007jh} which can be used to describe the process nonperturbatively at the quark-gluon level.
             Once the underlying propagators, vertices and bound-state amplitudes are determined
             from Dyson-Schwinger, Bethe-Salpeter and covariant Faddeev equations and satisfy the relevant Ward identities, the approach provides a well-defined
             construction principle for several reactions such as Compton scattering, pion-nucleon scattering, pion electroproduction or $\pi\pi$ scattering~\cite{Eichmann:2011ec}.
             A variety of meson properties, and more recently also nucleon and $\Delta$ elastic and transition form factors have already been calculated
             in that framework, see~\cite{Chang:2011vu,Eichmann:2011ej} for overviews.

  \section{Structure of the nucleon Compton scattering amplitude}

             Fig.~\ref{fig:qcv-born} shows the separation of the nucleon Compton amplitude into a Born contribution  and a one-particle irreducible (1PI) residual part,
             $\widetilde{J}^{\mu\nu} = J^{\mu\nu}_\text{B} + J^{\mu\nu}$.
             The former contains the nucleon poles in the $s-$ and $u-$channels and is given by
             \begin{equation}\label{nca-born}
                 J^{\mu\nu}_\text{B}(P,\Sigma,\Delta)  = -\Lambda_+^f \left[ \Gamma_N^\mu(-Q')\,S_N(P+\Sigma)\,\Gamma_N^\nu(Q)
                                                          +\Gamma_N^\nu(Q)\,S_N(P-\Sigma)\,\Gamma_N^\mu(-Q') \right] \Lambda_+^i \,.
             \end{equation}
             The residual part encodes the unknown structure information such as the nucleon's polarizabilities.
             The kinematics are illustrated in the figure: $\Delta$ is the four-momentum transfer,
             $\Sigma=\tfrac{1}{2}(Q+Q')$ is the average photon momentum, and $P=\tfrac{1}{2}(P_i+P_f)$
             is the average nucleon momentum. $S_N(p) = (-i\slashed{p}+M)/(p^2+M^2)$ is the nucleon propagator with nucleon mass $M$, and
             $\Lambda_+^{i,f} = \tfrac{1}{2}\left( \mathds{1}+\Slash{P}_{i,f}/iM\right)$ are the positive-energy projectors for the incoming and outgoing nucleon.
             The offshell nucleon-photon vertex is denoted by $\Gamma^\mu_N$; we suppressed its relative-momentum dependence for brevity.

     \begin{figure}[t]
     \center{
     \includegraphics[scale=0.25]{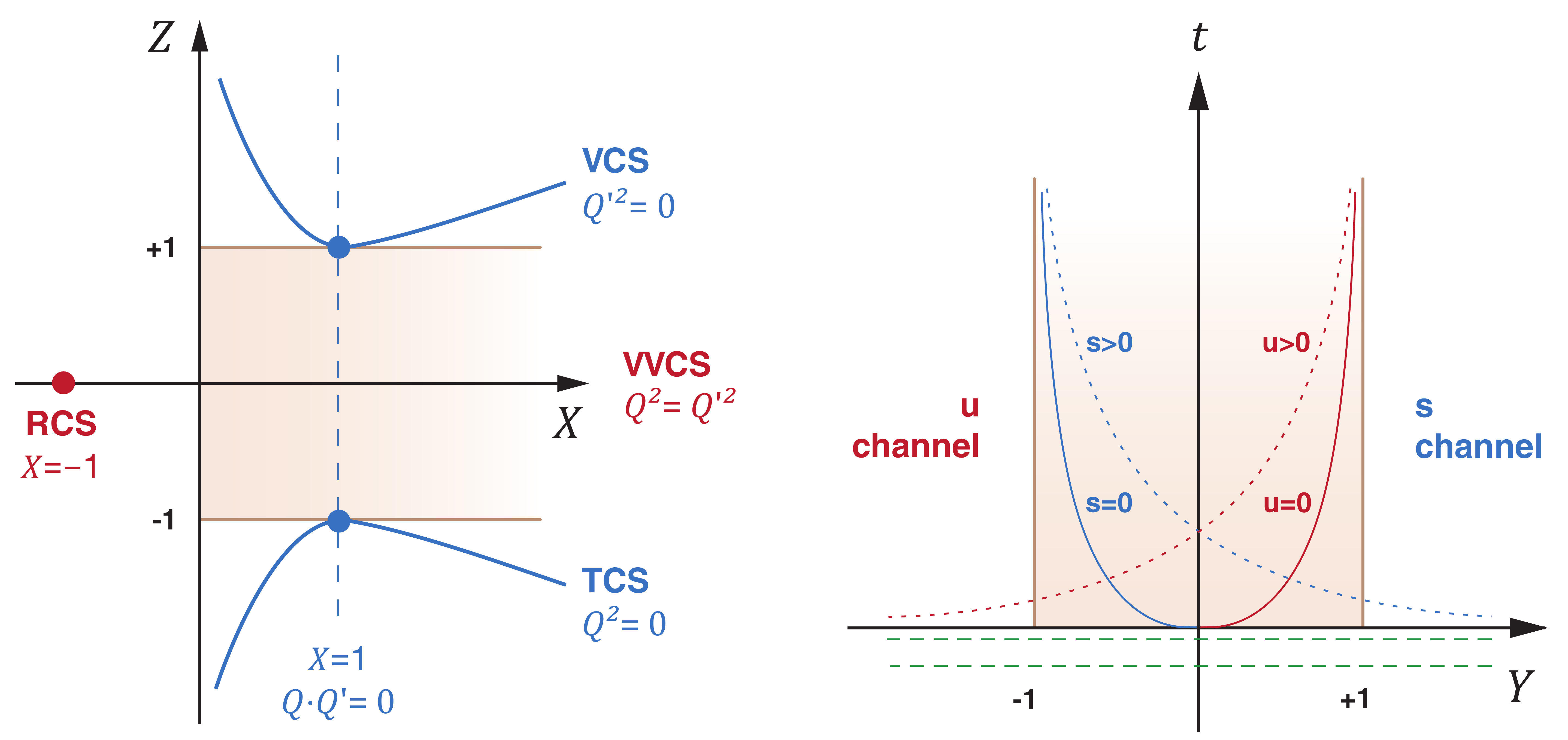}}
        \caption{\textit{Left panel:} $\{Q^2,{Q'}^2\}$ plane in the variables $X$ and $Z$.
        The abbreviations RCS, VCS, VVCS and TCS
        denote real, virtual, doubly virtual and timelike Compton scattering. The colored area describes the spacelike region.
        \textit{Right panel:} Mandelstam plane for real Compton scattering in the variables $t$ and $Y$.
        The physical $s-$channel region corresponds to $Y\geq 1$.
        The inserted lines visualize the resonance locations: the $t-$channel scalar and pion poles (dashed),
        $s-$ and $u-$channel nucleon poles from the Born terms (solid), and nucleon resonances (dotted).
        Our Mandelstam variables $s$ and $u$ are shifted by $-M^2$ compared to their usual definition, so that $s=0$ and $u=0$ correspond to the nucleon pole locations.}
        \label{fig:phasespace}
     \end{figure}

             Direct experimental information is available in the kinematic limits of real (RCS), virtual (VCS) and doubly-virtual forward Compton scattering (VVCS),
             where one or several Lorentz-invariant combinations of the photon momenta vanish, cf.~Fig.~\ref{fig:phasespace}.
             The structure of the scattering amplitude in these limits is closely tied to electromagnetic gauge invariance which
             entails that $\widetilde{J}^{\mu\nu}= J^{\mu\nu}_\text{B} + J^{\mu\nu}$
             is transverse with respect to the incoming and outgoing photon momenta. In general,
             $J^{\mu\nu}_\text{B}$ and $J^{\mu\nu}$ are not individually transverse since the intermediate nucleon in Eq.~\eqref{nca-born} is offshell, and
             the separation into a Born term and a structure part is in principle arbitrary as the former contains also an offshell nucleon-photon vertex.
             In that respect it is useful to employ the onshell Dirac form for the vertex with Dirac and Pauli form factors $F_1$ and $F_2$,
             \begin{equation}\label{dirac-current}
                -i \Gamma^\mu_N(Q) =  F_1(Q^2)\,\gamma^\mu + \frac{iF_2(Q^2)}{4M}\,[\gamma^\mu,\slashed{Q}]\,,
             \end{equation}
             since it guarantees that the Born term, and hence also the structure part, are transverse on their own
              even in offshell kinematics. This is usually not true for other onshell-equivalent forms of Eq.~\eqref{dirac-current}.
             Transversality and analyticity then imply that $J^{\mu\nu}$ is at least linear in both photon four-momenta $Q$ and $Q'$,
             whereas the Born term and its irregular low-energy limit is determined by experimentally known nucleon properties. 
             This is the essence of the low-energy theorem for Compton scattering, see Ref.~\cite{Scherer:1996ux} for a detailed discussion.

   \renewcommand{\arraystretch}{0.9}

             The structure part can now be decomposed in terms of 18 transverse tensor structures: 
             \begin{equation}\label{JR-basis}
                 J^{\mu\nu}(P,\Sigma,\Delta) = \sum_{i=1}^{18} F_i(t,X,Y,Z)\,\mathsf{T}^{\mu\nu}_i(P,\Sigma,\Delta),
             \end{equation}
             with coefficients that depend on four Lorentz-invariant kinematic variables.
             For example, one can work with the Mandelstam variable\footnote{Note that in Euclidean kinematics all scalar products pick up a minus sign  and that our definition
             of $t$ differs by a factor $-4M^2$ from the standard convention in the literature.}
             $t$, the incoming and outgoing photon virtualities,
             and the dimensionless photon crossing variable $\nu$:
             \begin{equation}
                 \frac{\Delta^2}{4M^2} = t\,, \quad
                 \frac{Q^2}{4M^2} = \tau\,, \quad
                 \frac{{Q'}^2}{4M^2} = \tau', \quad
                 \frac{\Sigma\cdot P}{M^2} = -\nu\,.
             \end{equation}
             We find it convenient to construct alternative variables from $P$, $\Sigma$ and $\Delta$:
             \begin{equation}\label{XYZ-def}
                 \frac{\Delta^2}{4M^2} = t\,, \quad
                 \frac{\Sigma^2}{M^2} =:  tX\,, \quad
                 \widehat{P}\cdot\widehat{\Sigma_T}  =: Y\,,  \quad
                 \widehat{\Sigma}\cdot\widehat{\Delta}  =: Z\,,
             \end{equation}
             where hats denote normalized four-momenta and the subscript $T$ refers to a transverse projection
             with respect to the momentum transfer $\Delta$; see Ref.~\cite{Eichmann:2012mp} for details. 
             Since the nucleon momenta are onshell, one has $P^2=-M^2(1+t)$ and $P\cdot\Delta=0$.

             The second choice is practical since the resulting phase space becomes quite simple, see Fig.~\ref{fig:phasespace}.
             The spacelike domain that enters two-photon exchange integrals with charged sources is the region $t$, $X>0$ and $Y$, $Z \in (-1,1)$.
             Real Compton scattering corresponds to $X=-1$ and $Z=0$,
             where the physical $s-$channel region is defined by $t \geq 0$ and $Y \geq 1$, with $t=0$ forward and $Y=1$ backward scattering.
             The generalized polarizabilities are defined in the limit $X=Z=1$ and $Y=0$, and forward VVCS corresponds to $t=0$ and $Z=0$.
             While the structure part $J^{\mu\nu}$ is analytic in these limits with respect to the nucleon poles,
             it still contains $s-$ and $u-$channel nucleon resonances, $t-$channel meson poles, $\rho-$meson poles at fixed $Q^2$ and ${Q'}^2$,
             and various cut structures from intermediate meson-exchange contributions.

             It is desirable to work with a tensor basis $\mathsf{T}^{\mu\nu}_i$ in Eq.~\eqref{JR-basis} that is free of kinematic singularities
             and implements the analyticity constraints per construction,  
             so that the resulting 18 dressing functions remain constant when either
             of the kinematic variables vanishes. Such bases have been constructed in the literature, see~\cite{Tarrach:1975tu,Drechsel:1997xv,Gorchtein:2009wz} and references therein,
             but the resulting expressions can become quite lengthy.
             In Ref.~\cite{Eichmann:2012mp} we have proposed an alternative, compact tensor basis which is regular in these kinematic limits.   
             The template for its construction is the simpler case of a fermion-photon vertex, where we define the quantities 
             \begin{equation} \label{new-transverse-projectors-2}
                   t_{ab}^{\mu\nu} := a\cdot b\,\delta^{\mu\nu} - b^\mu a^\nu\,,  \qquad
                   \varepsilon^{\mu\nu}_{ab} := \gamma_5\,\varepsilon^{\mu\nu\alpha\beta}a^\alpha b^\beta\,,
             \end{equation}
             with $a^\mu, b^\mu \in \{ P^\mu, \, Q^\mu,\,{Q'}^\mu \}$. Both expressions are regular in the limits $a\rightarrow 0$ or $b\rightarrow 0$, and
             $t_{ab}^{\mu\nu}$ is transverse to $a^\mu$ and $b^\nu$
             whereas $\varepsilon^{\mu\nu}_{ab}$ is transverse to $a$ and $b$ in both Lorentz indices.
             We combine them via
             \begin{equation}\label{EFG}
             \begin{split}
                 \mathsf{E}_\pm^{\mu\alpha,\beta\nu}(a,b) &:= \tfrac{1}{2}\left(\varepsilon^{\mu\alpha}_{Q'a'} \, \varepsilon^{\beta\nu}_{bQ} \pm \varepsilon^{\mu\alpha}_{Q'b'} \, \varepsilon^{\beta\nu}_{aQ} \right), \\
                 \mathsf{F}_\pm^{\mu\alpha,\beta\nu}(a,b) &:= \tfrac{1}{2}\left( t^{\mu\alpha}_{Q'a'} \, t^{\beta\nu}_{bQ} \pm t^{\mu\alpha}_{Q'b'} \, t^{\beta\nu}_{aQ} \right), \\
                 \mathsf{G}_\pm^{\mu\alpha,\beta\nu}(a,b) &:= \tfrac{1}{2}\left(\varepsilon^{\mu\alpha}_{Q'a'} \, t^{\beta\nu}_{bQ} \pm t^{\mu\alpha}_{Q'b'} \, \varepsilon^{\beta\nu}_{aQ} \right),
             \end{split}
             \end{equation}
             where primed momenta are obtained from $a = P,\, Q,\, Q'$ $\leftrightarrow$ $a' = P,\, Q',\, Q$.
             These tensors are now symmetric or antisymmetric under photon crossing and charge conjugation,
             transverse with respect to ${Q'}^\mu$ and $Q^\nu$, at least linear in both momenta, and free of kinematic singularities.
             They also have a simple Dirac structure as they are either proportional to $\mathds{1}$ or $\gamma_5$.
             We finally define the contractions
             \begin{equation}\label{EFG-Lambda}
             \begin{split}
                 \mathsf{E}_\pm^{\mu\nu}(a,b) &= \Lambda_+^f\,\mathsf{E}_\pm^{\mu\alpha,\alpha\nu}(a,b)\,\Lambda_+^i\,, \\
                 \widetilde{\mathsf{E}}_\pm^{\mu\nu}(a,b) &= \Lambda_+^f\,\mathsf{E}_\pm^{\mu\alpha,\beta\nu}(a,b)\,\tfrac{1}{6}\,\big[ \gamma^\alpha, \gamma^\beta, \slashed{P} \big]\, \Lambda_+^i\,,
             \end{split}
             \end{equation}
             with analogous expressions for $\mathsf{F}$ and $\mathsf{G}$,
             where we have used the totally antisymmetric combination of three $\gamma-$matrices, $\left[A,\,B,\,C\right] := \left[A,\,B\right] \,C + \left[B,\,C\right]\,A+\left[C,\,A\right]\,B$.

     \begin{table}[t]
     \center{
     \includegraphics[scale=0.89]{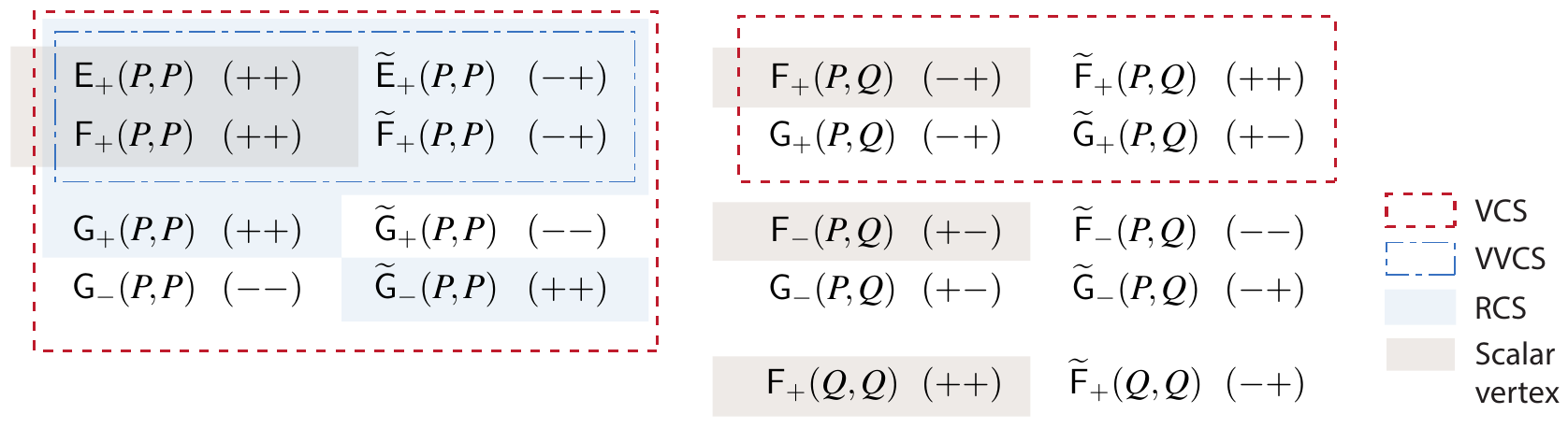}}
        \caption{18-dimensional, regular tensor basis for nucleon Compton scattering as defined in Eq.~(2.8); we suppressed the Lorentz indices $\mu\nu$ for better readability.
                 To ensure that all dressing functions
                 are even in the variables $Y$ and $Z$, the elements with $(-+)$, $(--)$ or $(+-)$ must be multiplied with $(P\cdot\Sigma) \sim Y$, $(\Sigma\cdot\Delta) \sim Z$, or
                 $(P\cdot\Sigma)\,(\Sigma\cdot\Delta) \sim YZ$, respectively.
                 The colored boxes show which elements survive in the various kinematic limits of RCS, VCS and forward VVCS; we also highlighted
                 the five elements that remain for a scalar Compton vertex.}
        \label{fig:tensorbasis}
     \end{table}

             A complete, linearly independent tensor basis free of kinematic singularities is then given by the 18 elements in Table~\ref{fig:tensorbasis}.
             Each basis element is symmetric or antisymmetric under photon crossing and/or charge conjugation which is indicated by the signs in the brackets
             (the first sign corresponds to photon crossing and the second to charge conjugation).
             Since the variable $Y$ from Eq.~\eqref{XYZ-def} is proportional to the crossing variable, it switches sign under photon crossing;
             the skewness variable $Z\sim \tau-\tau'$ flips its sign under both operations. By attaching appropriate combinations of $Y$ and $Z$ to the basis elements
             one ensures that all Lorentz-invariant dressing functions $F_i(t,X,Y,Z)$ are even in both $Y$ and $Z$.

             A real photon has only two physical polarizations, so that a contraction with the photon polarization vectors in the limits of VCS (${Q'}^2=0$)
             and RCS ($Q^2={Q'}^2=0$) eliminates certain tensor structures, or makes them linearly dependent upon each other, due to the prefactors $Q^2$ or ${Q'}^2$
             that appear in the quantities $t^{\mu\nu}_{QQ}$ and $t^{\mu\nu}_{Q'Q'}$ of Eq.~\eqref{new-transverse-projectors-2}.
             In the forward limit: $Q^\mu={Q'}^\mu$, so that several basis elements vanish or become redundant as well.
             In addition, the structures that are odd under charge conjugation disappear in RCS and forward VVCS as they are multiplied with a factor $Z=0$.
             This leaves 12 independent structures for VCS, six elements for RCS, and four independent elements in the case of forward VVCS, in agreement with previous analyses.
             $\mathsf{E}^{\mu\nu}_+(P,P)$ encodes the magnetic polarizability $\beta$ and $\mathsf{F}^{\mu\nu}_+(P,P)$ the electric polarizability $\alpha$.
             Table~\ref{fig:tensorbasis} also contains the five spin-independent structures that appear in the Compton amplitude of a scalar particle.

     \begin{figure*}[t]
     \center{
     \includegraphics[scale=0.094]{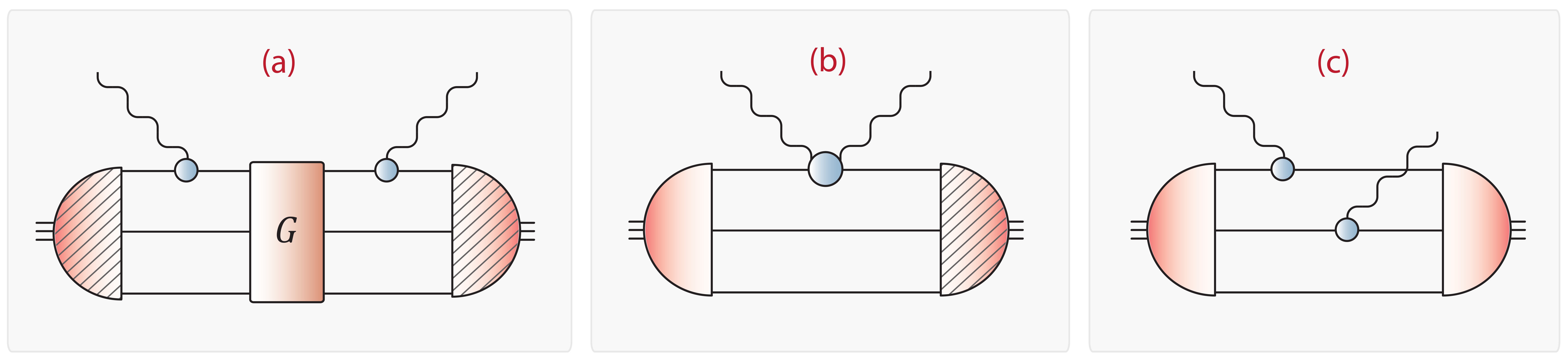}}
        \caption{Graphical representation of the nucleon Compton scattering amplitude in a rainbow-ladder truncation.
                 All propagators are dressed, and permutations in the quark lines and symmetrizations of the photon legs are not displayed.
                 The hatched amplitudes absorb two-body kernels that connect the spectator quarks.
                 Diagram (b) contains the 1PI part of the quark Compton vertex and diagram (a) provides the Born parts.}
        \label{fig:nca}
     \end{figure*}

  \section{Decomposition at the quark-gluon level}

             The actual structure of the residual part $J^{\mu\nu}$ of the nucleon Compton amplitude has to be determined dynamically.
             The decomposition of the total amplitude $\widetilde{J}^{\mu\nu}$ in the Dyson-Schwinger approach is visualized in Fig.~\ref{fig:nca} in the specific case of a rainbow-ladder truncation,
             which is our starting point for practical applications. Here the quark-quark interaction is simplified to a dressed gluon exchange,
             which is the basic (and only) parameter input in all calculations. 
             The resulting diagrams for Compton scattering can be arranged in three classes: diagram (a) contains the full six-quark Green function and
             thereby all nucleon resonances as well as handbag and cat's-ears contributions. Diagram (b) includes the 1PI quark
             two-photon vertex which also contributes to the handbag part (because the photons couple to the same quark), but also contains the full $t-$channel meson exchange structure.
             Diagram (c) provides further cat's-ears contributions.

             It is apparent that the earlier decomposition of Fig.~\ref{fig:qcv-born} loses its meaning at the quark-gluon level: any $s-$ or $u-$channel nucleon resonance,
             including the nucleon pole itself, is encoded in the quark six-point function, so that diagram (a) provides the nucleon Born terms as well as contributions from the structure part.
             The six-point function is also the biggest obstacle for a practical calculation; its self-consistent determination is beyond our present capacities.
             The sum of all (permuted, symmetrized) diagrams in Fig.~\ref{fig:nca} satisfies electromagnetic gauge invariance by construction and is therefore transverse.
             Unfortunately, this is not true for the individual diagrams which limits the usefulness of approximations. 
             In principle, one would need to extract the nucleon's polarizabilities from the full result, minus the Born diagrams from Eq.~\eqref{nca-born} at the nucleon level.

     \begin{figure*}[t]
     \center{
     \includegraphics[scale=0.31]{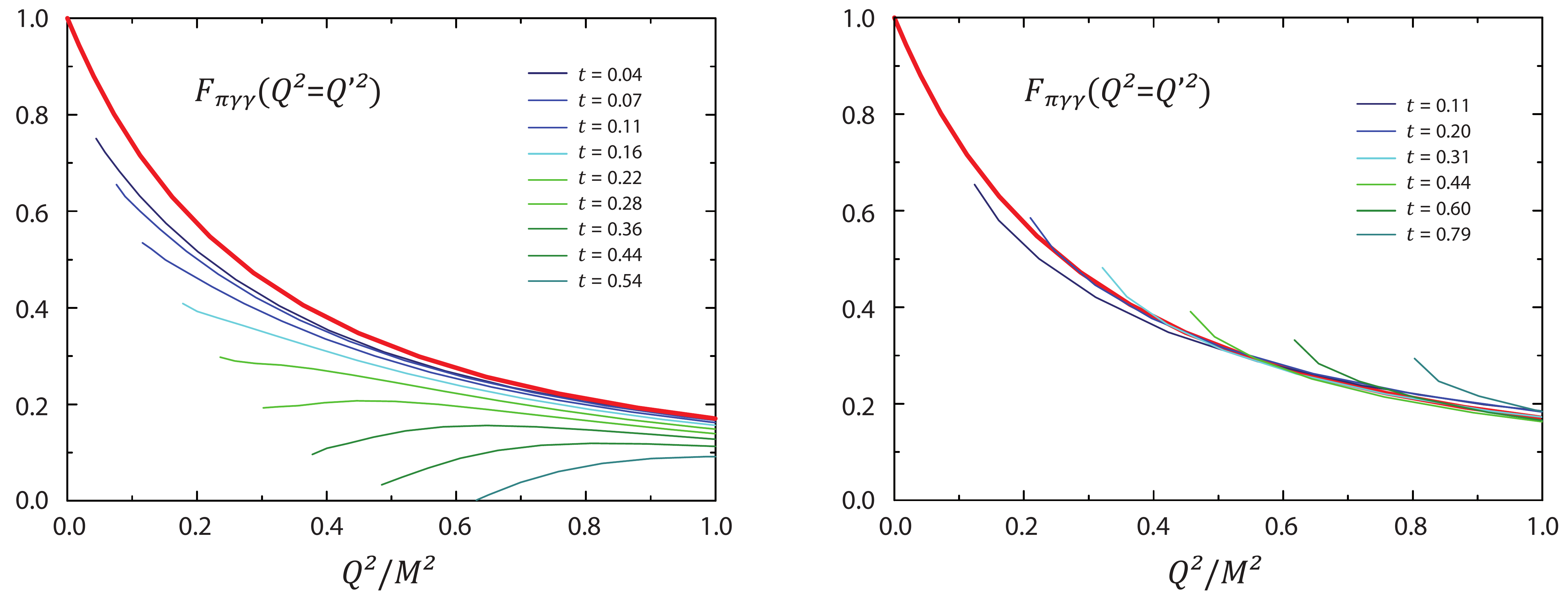}}
        \caption{Reconstruction of $F_{\pi\gamma\gamma}$ from the quark Compton vertex (\textit{left panel}) and the nucleon Compton amplitude (\textit{right panel}).
                 The thick red line in each panel is the result of the direct calculation for the $\pi^0\gamma\gamma$ transition current~\cite{Maris:2002mz};
                 the thin lines are the extrapolated values at various values of $t$.}
        \label{fig:results-2}
     \end{figure*}

             In Ref.~\cite{Eichmann:2012mp}  we have calculated the handbag content of the scattering amplitude. One can show that the sum of all handbag contributions
             from diagrams (a) and (b) constitute a quark Compton vertex which has the same form as in Fig.~\ref{fig:qcv-born}:
             $\widetilde{\Gamma}^{\mu\nu} = \Gamma^{\mu\nu}_\text{B} + \Gamma^{\mu\nu}$, with the generalization that the quark is now offshell
             and that the resulting structure of the vertex is considerably more complicated. In contrast to its analogue at the nucleon level,
             the Born part is now unambiguous as it depends on the dressed quark propagator and quark-photon vertex which we solve selfconsistently.
             Because the quark is offshell, the quark Compton vertex is no longer fully transverse. Instead, it has the following structure:
            \begin{equation}\label{qcv-wti-splitting-full}
                \widetilde{\Gamma}^{\mu\nu} =  \Gamma^{\mu\nu}_\text{B} + \Gamma^{\mu\nu}_\text{BC} + \Gamma^{\mu\nu}_\text{T} + \Gamma^{\mu\nu}_\text{TT}\,.
            \end{equation}
            Here, $\Gamma^{\mu\nu}_\text{BC}$ is completely determined by electromagnetic gauge invariance
            as it satisfies a Ward-Takahashi identity (WTI) that relates it to the dressed quark propagator. 
            $\Gamma^{\mu\nu}_\text{T}$ is also necessary to satisfy the WTI for the Compton vertex; it is partially transverse and determined by the transverse part of the quark-photon vertex.
            $\Gamma^{\mu\nu}_\text{TT}$ is a fully transverse remainder that is unconstrained. 

             The situation is analogous to the quark-photon vertex itself, whose WTI-preserving part is the Ball-Chiu vertex~\cite{Ball:1980ay} and determined by the quark propagator alone.
             When implemented in hadronic current matrix elements, it provides the dominant contribution to various electromagnetic form factors, 
             with $\rho-$meson poles (the 'vector-meson dominance part') stemming from the remaining transverse pieces.
             We have derived the generalization of the Ball-Chiu vertex to the two-photon case, $\Gamma^{\mu\nu}_\text{BC}+ \Gamma^{\mu\nu}_\text{T}$ from above, in Ref.~\cite{Eichmann:2012mp}.
             By implementing it in Fig.~\ref{fig:nca}~\!(b), it would in principle allow to test to what extent the handbag contribution to Compton scattering is determined by gauge invariance,
             apart from the further transverse part $\Gamma^{\mu\nu}_\text{TT}$ that encodes the $t-$channel resonances.

             If the result in Eq.~\eqref{qcv-wti-splitting-full} is applied to the Compton amplitude $\widetilde{J}^{\mu\nu}$, with the offshell quark replaced by an onshell nucleon,
             it provides also an interesting opportunity for model building at the nucleon level. Since $\widetilde{J}^{\mu\nu}$ and $J_\text{TT}^{\mu\nu}$ are both purely transverse,
             the combination $J^{\mu\nu}_\text{B} + J^{\mu\nu}_\text{BC}+ J^{\mu\nu}_\text{T}$ must be transverse as well. Thereby one obtains a gauge-invariant completion of the nucleon Born terms
             that incorporates the most general possible form of an offshell nucleon propagator and offshell nucleon-photon vertex
             beyond the Dirac form of Eq.~\eqref{dirac-current}.

             As discussed above, the quark Compton vertex $\widetilde{\Gamma}^{\mu\nu}$ is the microscopic representation of the nucleon handbag diagrams,
             in the same way as the quark-photon vertex determines the nucleon's electromagnetic form factors.
             In Ref.~\cite{Eichmann:2012mp} we have derived and numerically solved an inhomogeneous Bethe-Salpeter equation for the vertex
             that incorporates all four terms in Eq.~\eqref{qcv-wti-splitting-full}, i.e., Born, Ball-Chiu and transverse parts.
             The equation dynamically generates all $t-$channel meson poles which consequently appear in the Compton scattering amplitude.
             Their residues are the onshell meson two-photon transition currents, for example the $\pi^0\gamma\gamma$ transition form factor at the pion pole $t\rightarrow -m_\pi^2/(4M^2)$.
             Since the onshell currents are conserved, they are transverse and their extraction is unambiguous.
             Fig.~\ref{fig:results-2} shows that the calculated quark Compton vertex and nucleon Compton amplitude both reproduce the established result for the
             $\pi^0\gamma\gamma$ transition form factor in a rainbow-ladder truncation~\cite{Maris:2002mz}. 

             In absence of a complete dynamical calculation of the Compton amplitude $\widetilde{J}^{\mu\nu}$, the extraction of gauge-invariant polarizabilities is not possible without further assumptions.
             In principle, one may assume handbag dominance of the Compton scattering amplitude and augment the calculated handbag parts with kinematic terms to restore transversality at the nucleon level.
             This is our planned next step and would provide a nonperturbative model for the structure part in Fig.~\ref{fig:qcv-born}.
             Another application in the near future is the investigation of two-photon contributions to electromagnetic form factors
             that are determined from the nucleon Compton amplitude at spacelike momenta.

       \acknowledgments
       \small
     This work was supported by the Austrian Science Fund FWF under
     Erwin-Schr\"odinger-Stipendium No.~J3039,
            the Helmholtz International Center for FAIR
            within the LOEWE program of the State of Hesse,
            the Helmholtz Young Investigator Group No.~VH-NG-332, and by DFG TR16.

\end{document}